\documentclass[amsmath,longbibliography,superscriptaddress,reprint,showkeys]{revtex4-2}

\usepackage[T1]{fontenc}

\usepackage{amsmath,amsfonts,amssymb}

\usepackage{algorithm}
\usepackage{algorithmic}

\usepackage{array}

\usepackage{graphicx}
\usepackage{float}

\usepackage{textcomp}
\usepackage{verbatim}
\usepackage{url}

\usepackage{xcolor}
\usepackage{soul}
\usepackage[normalem]{ulem}

\usepackage{setspace}

\usepackage{chemformula}
\usepackage[version=4]{mhchem}

\usepackage{hyperref}
\usepackage{bookmark} 


\begin{document}

\title{Photonic Waveguide Circuit Integrated with Carbon Nanotube Single-Photon Source Operating at Room Temperature}

\author{Clement Deleau}
\affiliation{Quantum Optoelectronics Research Team, RIKEN Center for Advanced Photonics, Japan}

\author{Chee Fai Fong}
\affiliation{Quantum Optoelectronics Research Team, RIKEN Center for Advanced Photonics, Japan}
\affiliation{Nanoscale Quantum Photonics Laboratory, RIKEN Pioneering Research Institute, Japan}
\affiliation{Photonics-Electronics Integration Research Center, AIST, Japan}

\author{Finn L. Sebastian}
\affiliation{Institute for Physical Chemistry, Universität Heidelberg, Germany}

\author{Jana Zaumseil}
\affiliation{Institute for Physical Chemistry, Universität Heidelberg, Germany}

\author{Yuichiro K. Kato}
\affiliation{Quantum Optoelectronics Research Team, RIKEN Center for Advanced Photonics, Japan}
\affiliation{Nanoscale Quantum Photonics Laboratory, RIKEN Pioneering Research Institute, Japan}
\email{yuichiro.kato@riken.jp}


\begin{abstract}
Photonic integrated circuits require robust room-temperature single-photon sources to enable scalable quantum technologies. Single-walled carbon nanotubes (CNTs), with their unique excitonic properties and chemical tunability, are attractive candidates, but their integration into photonic circuits remains challenging. In this work, we demonstrate the integration of functionalized CNTs as room-temperature single-photon emitters into photonic cavities and waveguide circuits. (6,5) CNTs with aryl sp$^3$ defects are either stochastically deposited via drop-casting or deterministically positioned on photonic cavities using an anthracene-assisted transfer method guided by real-time photoluminescence monitoring. Photoluminescence spectra reveal cavity-enhanced emission, while second-order autocorrelation measurements confirm single-photon propagation through the photonic integrated circuit, highlighting the potential of CNTs for scalable, room-temperature quantum photonic applications.
\end{abstract}

\keywords{
Carbon Nanotubes, Photonic Integrated Circuits, Single Photon Emitter, Quantum Applications, Room Temperature, Photon Autocorrelation, Photonic Cavity, Waveguides.
}


\maketitle







\section{Introduction}

P\lowercase{h}otonic integrated circuits (PICs) are at the forefront of numerous emerging technologies, offering mass-manufacturable, stable, and compact optical solutions for a wide range of applications \cite{PICapplications,Chrostowski_PIC}. Environmental monitoring and telecommunications systems involving PICs are already commercially available \cite{PICapplications} and there is strong evidence that integrated quantum photonics applications will soon follow the same path \cite{PICquantumapplications,QuantumSiliconPhotonics}. Realizing such quantum circuits, however, requires efficient on-chip single-photon sources, whose scalable integration at room temperature remains a major challenge.

Epitaxial quantum dots offer high-performance emission properties but suffer from random spatial growth, are strongly affected by thermal quenching at room temperature and necessitate complex heterogeneous integration due to incompatibility with many PIC platform materials~\cite{Senellart2017, ScalableIntegration2022}. Colloidal quantum dots, though more easily incorporated into devices owing to their solution processability, are limited by unstable emission, photobleaching, and moderate single-photon purity~\cite{InPColloidalQDs2023}. In the past few years, more original approaches have been proposed. Native silicon nitride emitters, fully compatible with the SiN photonic platform, have been demonstrated to show single-photon generation at room temperature but are also affected by spontaneous—and thus random—emitter placement, and their emission mechanisms remain poorly understood~\cite{Senichev2021, Senichev2022}. Very recently, defect-based emitters in hBN flakes transferred onto waveguides have been introduced via laser writing, demonstrating a relatively easy way to deterministically implement ambient-condition emitters, but still face challenges including low single-photon purity and emission outside the telecom band \cite{yamashita2025deterministic, Abidi2019}.

Due to their one-dimensional structure and exceptionally stable excitonic states, single-walled carbon nanotubes (CNTs) have emerged as promising single-photon emitters for integrated photonics. They offer remarkable photostability, potentially resisting photobleaching and blinking effects, and tunable emission across the telecom bands (1100–1600~nm) via chirality selection and defect engineering~\cite{CNTsingleemitter2}.
In particular, introducing exciton trapping sites through molecular defect engineering enables CNTs to emit single photons with high purity at room temperature~\cite{CNTsingmeemitter1,CNTsingleemitter2}. Pristine CNTs have also demonstrated electrically driven photoluminescence (PL)~\cite{Mueller2010, HigashideKato} and have been integrated into photonic cavity structures~\cite{Pyatkov2016KrupkeCNTcavityWaveguide, Kato_PHC_CNT,CNTcavitycouplingMiurakato} as well as quantum photonic circuits operating at cryogenic temperatures~\cite{Khasminskaya2016_krupkeCNT_circuit}. These attributes establish CNTs as promising candidates for scalable quantum photonic applications; however, their implementation into functional on-chip circuits at room temperature remains challenging due to the stringent requirements for generating, coupling, and guiding single photons.

\begin{figure*}[t]
    \centering\includegraphics[width=0.95\textwidth]{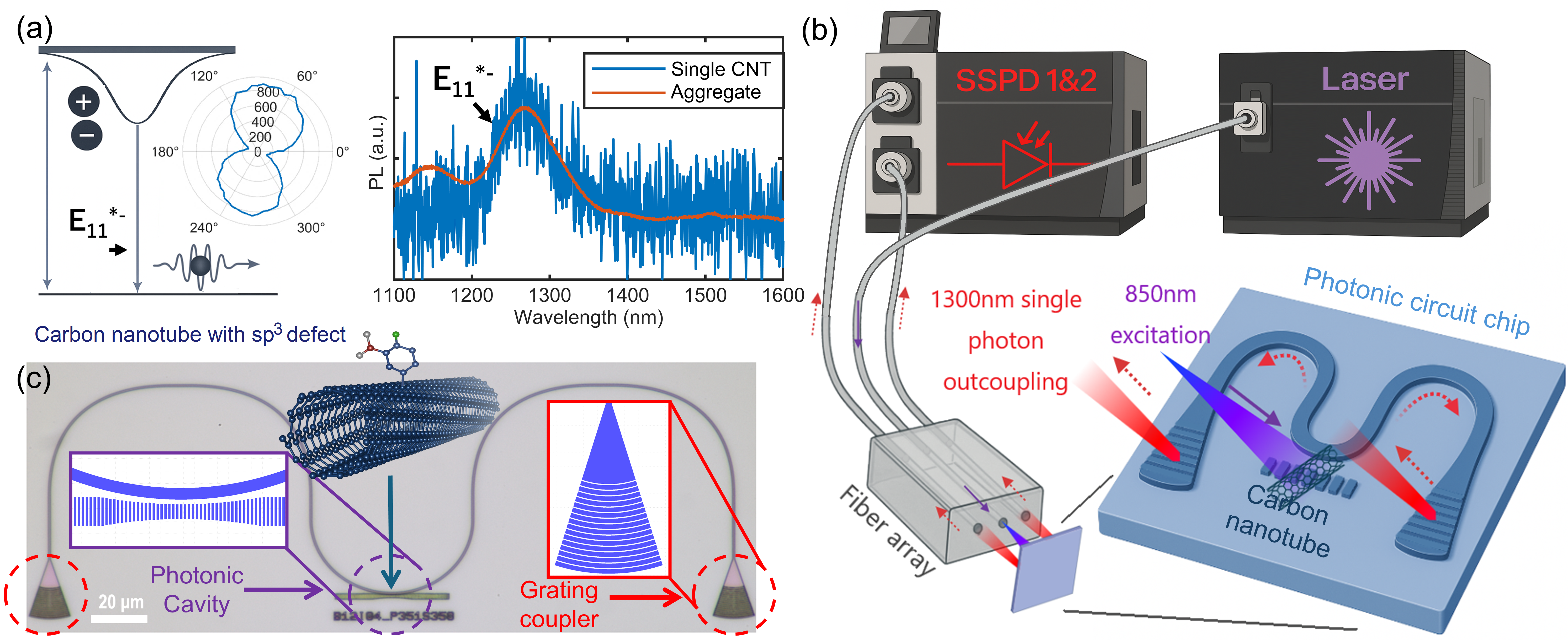}
    \caption{(a) Energy-level diagram of an exciton trapped in a CNT defect state, normalized emission spectra of a single (6,5) CNT emitter and a CNT aggregate after substrate solution coating, and polarization excitation dependence of a single CNT emitter. (b) Proposed optical setup schematic for the demonstration of CNT-emitted single-photon cavity coupling and waveguide propagation using a fiber array, a visible laser excitation, and superconducting single-photon detectors. (c) Optical microscope image of the fabricated base-design circuit, with cavity and grating layout schematics. The CNT image was generated with \cite{tubegen}.}
    \label{fig:fig1}
\end{figure*}

The objective of this research is to provide a platform for CNT-based quantum PIC applications by introducing strategies for both stochastic and deterministic cavity integration of individual CNTs and demonstrating their efficient coupling into PICs at room temperature. We begin by presenting the functionalization process and preparation of single-walled (6,5) CNTs, and outline their PL properties, as shown in Figure~\ref{fig:fig1}a in the context of subsequent implementation into PICs.
We then focus on the development of PICs, detailing the cleanroom fabrication process of a low-fluorescence lithium niobate (LN) waveguide platform, readily available for future electro-optical modulation. This is followed by the description of the photonic design and experimental characterization of waveguides, couplers, and cavities, measured with a custom automated testing setup. The same platform is subsequently integrated into a PL measurement system and adapted for CNT-circuit coupling analysis.
Using a fiber-coupling configuration depicted in Figure~\ref{fig:fig1}b, we demonstrate efficient coupling of CNT emission into single-resonant waveguide-coupled photonic cavities (Figure~\ref{fig:fig1}c), with enhanced and spectrally sharpened PL. Finally, we report single-photon emission, waveguide propagation and path routing at room temperature, confirmed through autocorrelation measurements. This work represents the first demonstration of a PIC that leverages CNT single-photon emission in ambient conditions, marking a key step toward realizing CNT-based room-temperature quantum applications.

\section{Carbon nanotube functionalization and optical characterization}
Owing to their widespread availability among CNT species and their well-characterized excitonic properties including spectral features, lifetimes, polarization, brightness, and autocorrelation, (6,5) single-walled CNTs are selected for functionalization and integration into photonic circuits.
In this work, we first functionalize chirality sorted (6,5) CNTs with aryl sp$^3$ defects in toluene. PFO-BPy-wrapped (6,5) SWCNTs are treated with 2-iodoaniline in the presence of KOtBu as a strong organic base, selectively generating defects with the $E_{11}^{*-}$ binding configuration~\cite{Finn1CNT65}, at low densities of approximately 1--2 defects~$\mu\mathrm{m}^{-1}$~\cite{Finn2CNT65}. 

Following functionalization, the CNTs are transferred into aqueous dispersions via surfactant-assisted processing. This strategy allows the coating of polymer substrates used in our integration technique, which are sensitive to most organic solvents, while preserving the optical properties of the CNTs. The transfer is achieved by filtration onto polytetrafluoroethylene membranes (Merck Millipore JVWP, 0.1~$\mu$m pore size), followed by repeated washing with hot toluene (6~$\times$~1~h, 80$^\circ$C), and redispersion via bath sonication (1.5~h) in an aqueous solution of 1\% (w/v) sodium deoxycholate (DOC, Sigma-Aldrich, $>$98.0\%). After centrifugation (2~$\times$~45~min, 60,000~$\times$~$g$), the supernatant is collected and diluted 50-fold.

PL measurements are then carried out using a homebuilt free-space setup comprising of optical components, a Princeton Instruments Acton 2300i spectrophotometer, and a Coherent Mira 900 laser operating at 850~nm in continuous-wave mode with 1~$\mu$W excitation power. Figure~\ref{fig:fig1}a displays the emission spectra of both an individual functionalized (6,5) CNT emitter with single defect and a CNT aggregate, and shows the excitation polarization dependence of an individual CNT. The $E_{11}^{*-}$ emission peak of these functionalized CNTs is centered around 1250--1300~nm, aligning with both the transparency window of most photonic circuit materials and the primary telecommunication wavelength range that benefits from zero-dispersion optical fibers.

\section{Photonic integrated circuit technology development}
Several different material platforms have been proposed for quantum photonics applications~\cite{PICquantumapplications,QuantumSiliconPhotonics}, each yielding specific performance profiles in key circuit characteristics such as component density, propagation losses or active optical components and intrinsic photon-source availabilities, among others. After evaluating several materials, we have selected LN for the development of our circuit technology, as it offers both high optical modulation efficiency and exceptionally low emission in the near-infrared range; two properties critical for applications such as quantum communication \cite{LNquantumphotonics} and for addressing the single photons from functionalized CNTs. The low emission of LN, in particular, is supported by comparative measurements of various thin films on insulator substrates, whose PL spectra in the near-infrared range are provided in Figure~\ref{fig:PLmaterial}.

\begin{figure*}[t]
    \centering\includegraphics[width=0.49\textwidth]{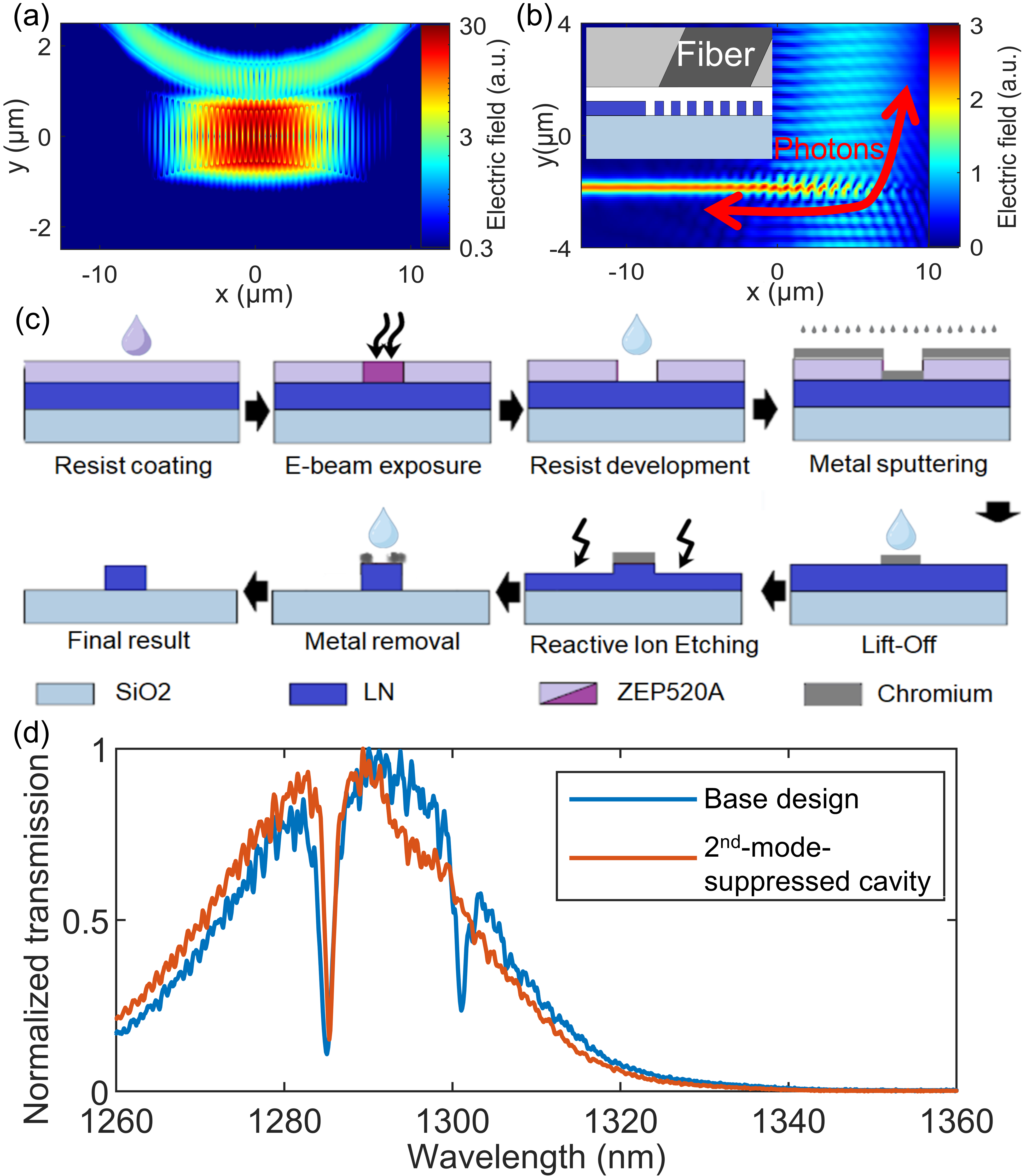}
    \caption{(a) Top-view FDTD-simulated optical field distribution of the designed cavities at a wavelength near resonance, (b) simulated optical field of the GC at the coupling wavelength, (c) Fabrication process of the LN PIC, and (d) measured transmission spectra for a base-design photonic cavity (blue) and a second-mode-suppressed cavity (orange).}
    \label{fig:Cavityspectraltransmission}
\end{figure*}

Grating couplers (GCs), waveguides, and cavities are first simulated in Lumerical, and hundreds of layout variants with diverse geometries are generated to comprehensively address fabrication and material tolerances (Supporting Information, Section~\ref{layoutsection}). In particular, grating photonic cavities are simulated using 2.5D finite-difference time-domain method and tailored to exhibit resonances around the CNT emission peak between 1260 and 1310~nm. Figure~\ref{fig:Cavityspectraltransmission}a,b respectively show the simulated field profiles of cavities and grating couplers.

Circuit fabrication follows a standard process illustrated in Figure~\ref{fig:Cavityspectraltransmission}c. Thin films of 300~nm LiNbO$_3$ on oxide wafers, purchased from Innosemitech, are initially coated with Zeon ZEP520A resist, and waveguide circuits are patterned using a high-resolution Advantest F7000S-VD02 electron-beam lithography system. After development in Zeon ZED-N50 solution and rinsing in water, a 70~nm thick chromium layer is sputtered onto the chip, with the thickness verified by profilometry. Chromium serves as a hard mask, as LN exhibits high resistance to dry etching resulting in low etch selectivity to organic resists. Following chromium deposition, a lift-off process is performed using sonication in dimethylacetamide solvent. Dry etching is then carried out using alternating cycles of SF$_6$ plasma etching and C$_4$F$_8$ passivation, operating for 55~minutes at a rate of 30~cycles per minute. The resulting LN etch depth is measured to be 240~nm, forming rib-type waveguides. Residual chromium is subsequently removed by immersion in a chromium etchant solution with sonication for 20~minutes. While many PIC platforms typically cover waveguides with oxide or organic layers, here the waveguides are left exposed for CNT coupling after deposition.

For circuit testing, a characterization platform is established using fiber V-groove arrays, a Santec TSL-550 tunable laser centered around 1300 nm, a photodiode, and a linear translation stage controlled by custom LabVIEW algorithms (Figure~\ref{fig:fig1}b). The experimental setup is shown in Supporting Information, Section~\ref{setupsectionsupporting}. Fiber arrays (FAs), which align multiple optical fibers with high precision, are commonly used to enable efficient coupling to photonic chips. Light is injected and collected via integrated LN GCs at the ends of the waveguides, enabling diffraction-based out-of-plane optical coupling. This configuration allows the FA to simply hover above the chip surface for effective PIC characterization. By matching the GC layout to the FA geometry, the design supports simultaneous multi-channel coupling. Measurement routines are fully automated via LabVIEW software, which interfaces directly with the design layout.

Grating coupling efficiency at 1280~nm is found to be approximately –9~dB per coupler, with a 70~nm bandwidth, exceeding the $E_{11}^{*-}$ emission peak. Although this coupling loss is substantial, significant improvement (down to 1.42~dB) could potentially be achieved through the implementation of cladding layers, bottom reflectors, subwavelength gratings, and chirped profiles, as reported in Ref. \cite{LNGcouplers}.

Waveguide propagation losses are estimated at approximately 3~dB/cm. Further improvements are feasible through optimization of the metal mask deposition and etching processes, with recent demonstrations achieving losses as low as 0.4~dB/cm for LN waveguides \cite{LNwaveguide}. The optimized cavity structures have a period of around 350~nm, a duty cycle of 0.55, and a grating width that quadratically tapers from 2.30~$\mu$m to 1.15~$\mu$m over a 10~$\mu$m cavity length. This relatively large cavity size is chosen to facilitate CNT coupling but leads to the appearance of a second longitudinal cavity mode (Figure~\ref{fig:Cavityspectraltransmission}d, blue curve). To suppress this unwanted mode, the periodicity of the external gratings is slightly reduced by 5$~$nm. This adjustment not only results in the clear elimination of the second mode but also leads to a slight enhancement of the resonance optical quality factor, reaching values up to 1200, as observed in the measured waveguide transmission shown as the orange curve in Figure~\ref{fig:Cavityspectraltransmission}d.

\section{Carbon nanotube circuit integration}

We explore two methods for incorporating CNTs onto photonic chips: direct drop-casting of CNT aqueous solution and anthracene-assisted transfer technique \cite{Anthracenetechnique1Fang2023, AnthraceneTransferKeigoNaturecoms}. While drop-casting provides a quick approach for preliminary tests, the uncontrolled CNT positioning limits its scalability. In contrast, the anthracene-assisted method, referred to here as CNT stamping, enables clean, deterministic placement of CNTs onto photonic cavities.

\begin{figure*}[t]
    \centering\includegraphics[width=0.9\textwidth]{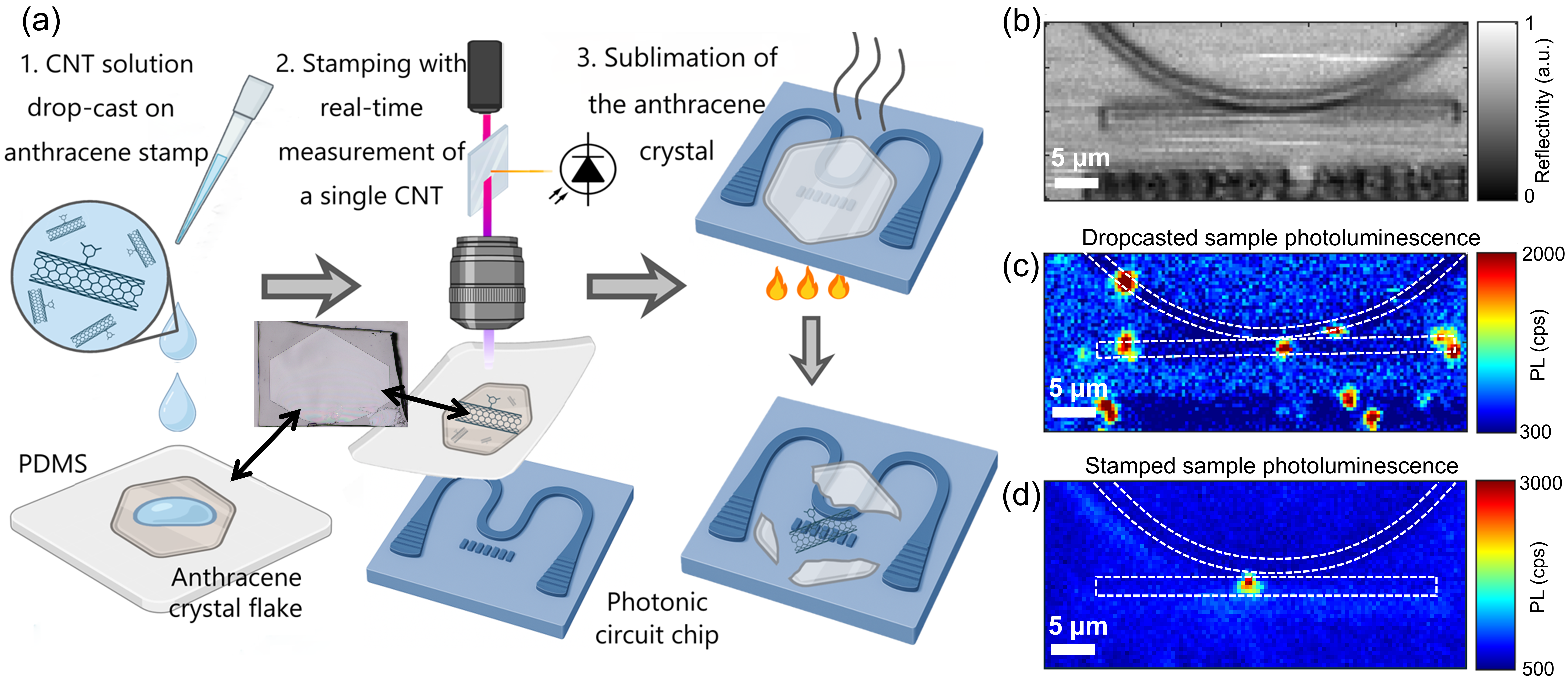}
    \caption{(a) Diagram for anthracene-assisted transfer technique and (b) cavity reflection mapping and PL mappings of (c) dropcasted cavity and (d) stamped cavity samples, with 780nm-1.5 $\mu$W of excitation.}
    \label{fig:fig3}
\end{figure*}

The CNT stamping process consists of three main steps, as illustrated in Figure~\ref{fig:fig3}(a). A 4500~$\mu$m thick Gel-Pak polydimethylsiloxane (PDMS) sheet is first cut into a 1~mm$^2$ piece and placed at the center of a glass sample holder. A custom-grown anthracene crystal flake is then carefully positioned on the PDMS square. To simplify the original multistep method \cite{AnthraceneTransferKeigoNaturecoms}, we directly coat the flake with water-based CNT dispersion, as anthracene is otherwise soluble in many organic solvents. While the PL quantum yield of CNTs is lower in water compared to toluene as dispersion solvent \cite{Finn3CNT,Finn2CNT65}, the DOC surfactant mitigates quenching and enables sufficient individualization for transfer. After 30 seconds, the droplet is removed via capillary action, leaving the flake sparsely coated with functionalized (6,5) CNTs and the stamp sample is subsequently mounted on a motorized translation stage. Detailed mapping images and setup photographs are given in the Supporting Information, Section~\ref{sectionstampingimage}.

In the second step, the CNTs are located by PL mapping through the transparent PDMS and glass using a continuous-wave excitation beam that is vertically polarized during mapping to preferentially locate CNTs aligned with the cavity mode polarization, thus avoiding the need to rotate the sample prior to stamping. Although photon autocorrelation measurements can be used to assess the single-photon emission characteristics of the CNTs, they are typically avoided due to photobleaching risks associated with prolonged pulsed excitation. Instead, rapid lifetime measurements are favored, offering a qualitative indication of single quantum emitter presence which tend to give longer decay times than CNT aggregate, as demonstrated in \cite{CNTdecaytimesinglebundle}. Once a suitable CNT is identified, a second translation stage holding the PIC chip aligns the cavity with the excitation beam as a visual reference using a microscope-mounted camera aligned with the collection/excitation path, ensuring that both the CNT and the target cavity are centered along the optical axis. Contact is made by gradually advancing the PIC toward the stamp while monitoring PL to confirm alignment. Due to inherent non-planarity in the PDMS surface,  fine positional adjustments guided by changes in CNT PL intensity, typically on the order of a few micrometers, are often needed during approach. Afterward, the system is held stationary for 30 minutes before carefully separating the stamp. CNT transfer may occur with or without the anthracene flake, and several attempts are typically required. Substrate adhesion properties significantly influence the stamping yield. In most cases, only partial stamping occurs even after multiple attempts. While CNT positioning is generally straightforward, PDMS retraction remains the most delicate step and is currently the primary bottleneck of the method, with an overall success rate of~25$\%$. Surface treatment strategies are being explored to improve adhesion and reproducibility. 

The third step involves heating the chip to 80$^\circ$C via a thermoelectric cooler to gradually sublimate the anthracene, leaving the CNT cleanly positioned on the cavity. Figure~\ref{fig:fig3}b-d presents representative PL and reflectivity maps of coupled cavity circuits prepared by both drop-casting and stamping, and additional stamp mapping results are shown in the Supporting Information, Section~\ref{sectionstampingimage}.

\section{Results}
After placing the CNT on the chip, we demonstrate efficient CNT coupling to the waveguide circuit using our FA-based characterization platform. As illustrated in Figure~\ref{fig:fig1}b, three consecutive FA channels are employed: the central channel for CNT/cavity excitation, and the two outer channels are connected to the GCs. This way, CNT photons emitted in the cavity are coupled to the waveguide circuit propagating towards either GCs and subsequently transferred to the external FA channels. This configuration simplifies alignment and testing, and inherently provides a 50/50 splitter for autocorrelation measurements due to the symmetry of the coupling circuit around the photonic cavity. Because of the large mode field diameter of the fiber and the lack of focusing optics at the output, the illuminated area on the chip is relatively wide, with roughly ~1\% of the injected power overlapping with the cavity mode, necessitating higher excitation power.

\begin{figure*}
    \centering\includegraphics[width=0.49\textwidth]{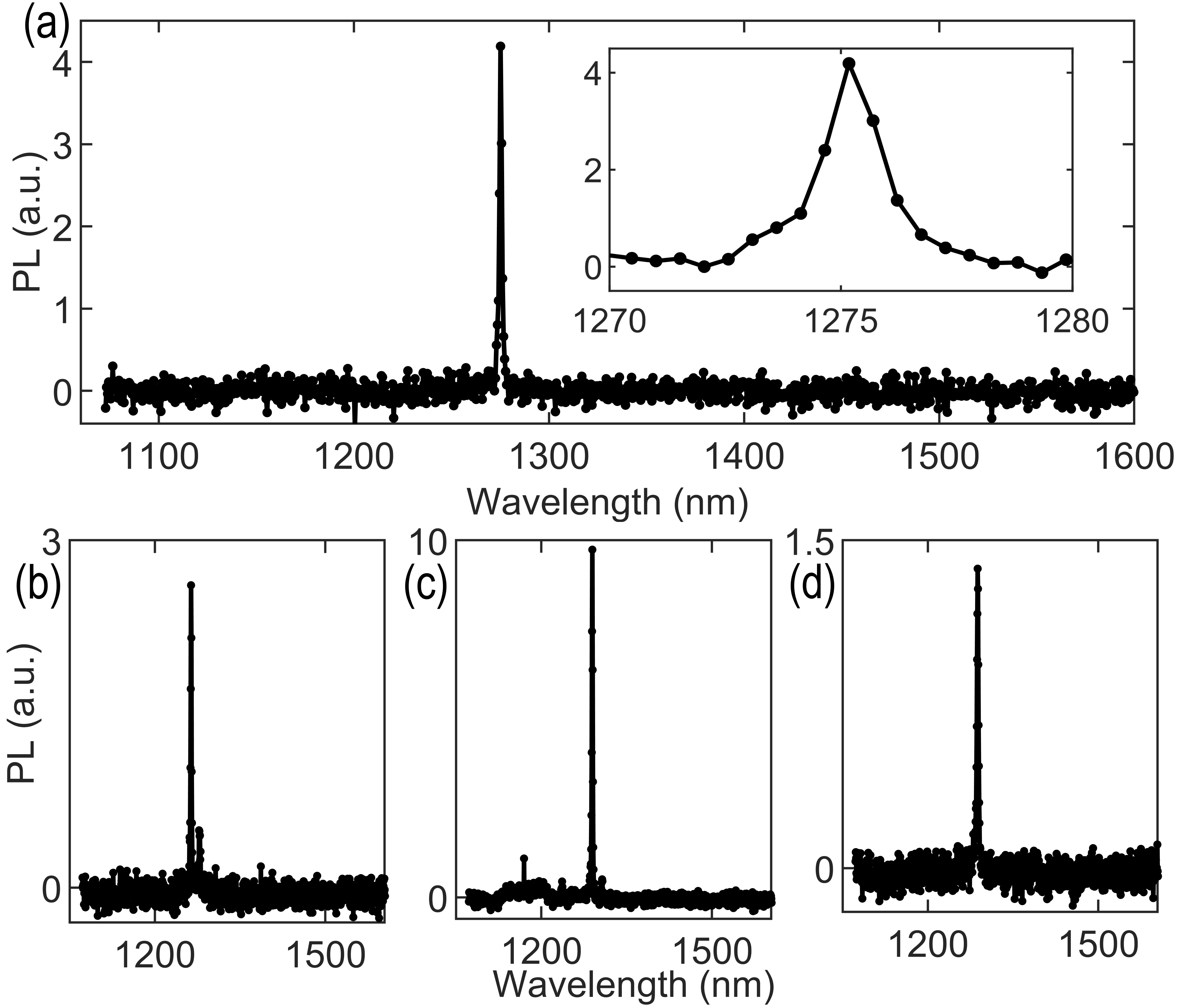}
    \caption{Cavity-coupled CNT PL spectra with 780 nm-1.5 $\mu$W of excitation. (c) and (d) are not confirmed to be single-photons emitters.}
    \label{fig:bestresospectrum}
\end{figure*}

\begin{figure*}
    \centering\includegraphics[width=0.49\textwidth]{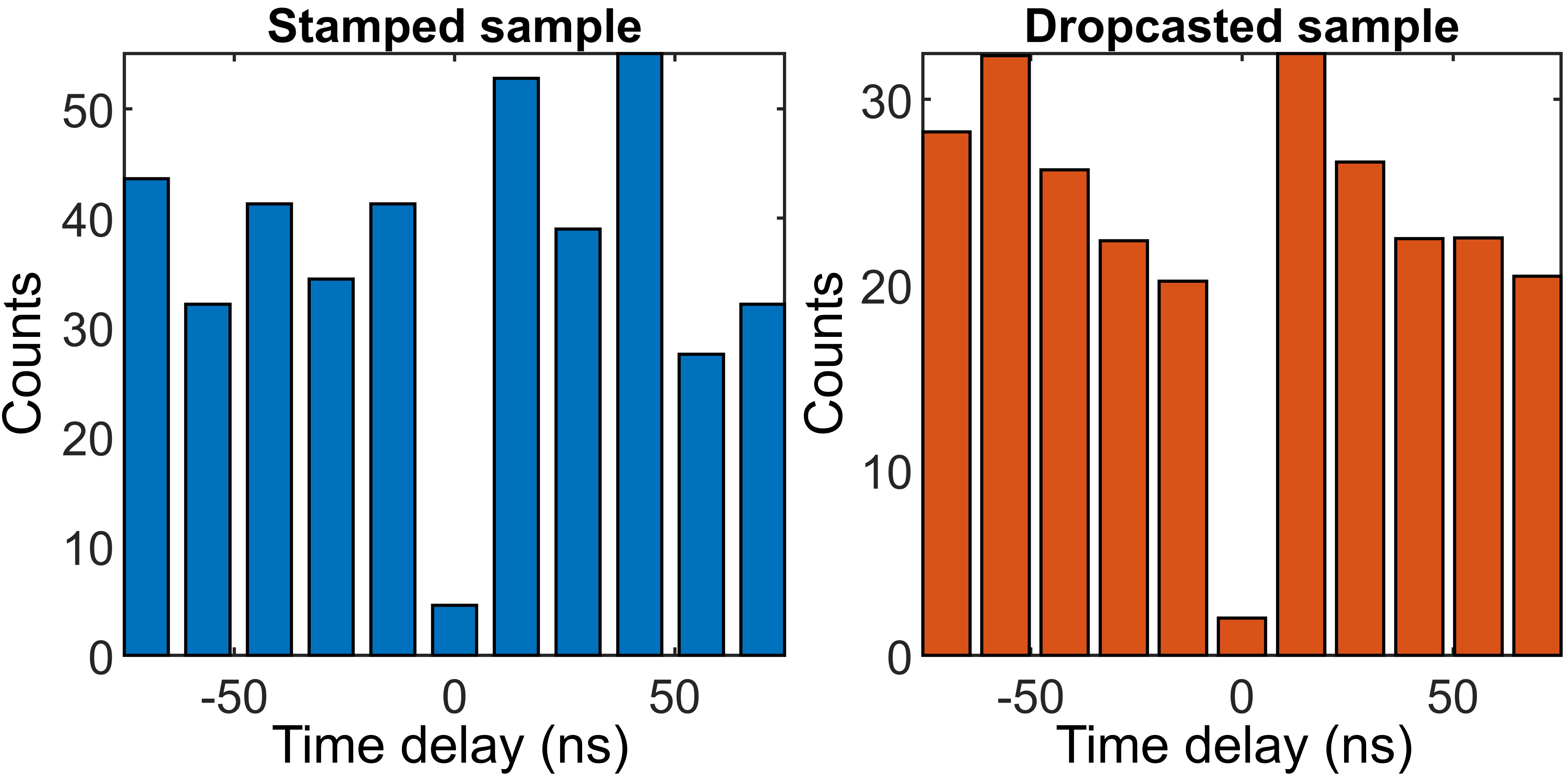}
    \caption{Measured autocorrelation results of the best stamped and dropcasted samples.}
    \label{fig:autocorrel}
\end{figure*}
PL coupling spectra are first collected by connecting either of the outer FA channels to a spectrometer. A sharp emission peak may appear at the location of the resonance dip in the transmission spectrum, as shown in Figure~\ref{fig:bestresospectrum}. A slight degradation of the resonance quality factor is observed, which we attribute to residual polymer from the CNT wrapping, increasing the internal cavity losses. Depending on CNT positioning or the possible presence of additional CNTs, different spectral profiles may be observed, as described in the Supporting Information, Section~\ref{sectionupportingcavitycoupledspectra}.

To assess single-photon emission from the waveguided photons, each of the two outer FA channels is connected to a superconducting single-photon detector from Single Quantum, which is in turn connected to a Qutools quTag module for time-correlated single-photon counting and event tagging. To suppress residual excitation light coupled into the waveguide, each fiber line includes a 1200~nm long-pass filter mounted between a pair of collimators. Due to the overall low PL quantum yields of CNTs, continuous-wave autocorrelation measurements are slow. Although the pulsed lasing mode can gradually damage the CNTs, the laser is switched to a 76~MHz mode-locked excitation to accelerate data acquisition, making it suitable primarily for proof-of-concept experiments at low brightness. An increase in blinking behavior is also observed under pulsed excitation, which can also serve as an indicator of single-emitter characteristics, despite the pulsed mode's adverse effects on long-term emitter stability.

Autocorrelation measurements yield \( g^{(2)}(0) \) values of \( 0.08 \pm 0.06 \) and \( 0.10 \pm 0.06 \) for the best-performing stamped and drop-cast samples (Figure~\ref{fig:autocorrel}), demonstrating clean emission and successful waveguide propagation of single photons at room temperature. On average, drop-cast samples tend to show lower single-photon purity, likely due to the greater difficulty in isolating a single coupled CNT from surrounding individual tubes within the excitation area near the waveguide. It should also be noted that the total number of tagged coincidence events in most measurements remains relatively low, primarily because the CNTs degrade or cease emitting after minutes to a few hours under continuous pulsed excitation. In the Supporting Information, Section~\ref{sectionsupportingstability}, we present multiple examples of emitter blinking and degradation behavior. We do not attribute these effects directly to the presence of water molecules, as demonstrated by measurements in which the CNT environment is filled with $N_{2}$.

The PL decay time of the CNTs remains essentially unchanged with cavity coupling, as shown in the Supporting Information, Section~\ref{sectionsupportinglifetime}, contrary to the shortening typically expected from Purcell enhancement. This can be attributed to the cavity mode, which both limits overlap with CNTs in the grating gaps due to its dielectric nature and is designed with a relatively large mode volume (1.16~$\lambda^3/n^3$, 0.52~$\mu$m$^3$, where $\lambda$ is the wavelength and $n=1.71$ is the effective refractive index) to facilitate CNT stamping.
The Purcell factor for broad linewidth emitters can be expressed as
\[
F_\mathrm{P} = \frac{3 \lambda^3 Q_e}{4 \pi^2 n^3 V} \, \xi,
\]
where $Q_e$ is the emitter quality factor, $V$ the cavity mode volume, and $\xi$ accounts for spatial and polarization overlap \cite{Purcellfactorbroademitter}. For our coupled CNTs, this yields a maximal $F_\mathrm{P}$$\sim$0.34 assuming a mode overlap of $\sim$30\% for a CNT in an airgap. Combined with the low radiative efficiency of functionalized (6,5) CNTs on oxide substrates ($\sim$2.4\%) \cite{AkihiroCNTphotonicrystalfunctionalized}, this explains the negligible enhancement of decay time.

Considering the measured free-space PL of uncoupled CNTs collected with our 0.65~NA objective, and correcting for waveguide collection efficiency ($-9$dB) and the maximal Purcell factor, we conservatively estimate that $\sim$30\% of CNT emission is coupled into the waveguide circuit. Potential performance improvements include using air-mode cavities with smaller mode volumes (as low as 0.0237$\lambda^3/n^3$ \cite{CNTcavitycouplingMiurakato}) and improved spatial overlap with the CNT, which could respectively enhance the Purcell factor by $\sim$50 times and $\sim$3 times. Together, these modifications could potentially boost the total emission rate by $\sim$150 times compared to the current design.


\section{Conclusion}
In this work, we have demonstrated cavity-enhanced emission and single-photon coupling of functionalized CNTs into a PIC at room temperature. Individual CNTs are integrated onto the chip using two approaches: a stochastic solution-based dropcasting method and a more deterministic anthracene-assisted transfer technique. In both cases, second-order autocorrelation measurements confirmed single-photon emission and propagation, with $g^{(2)}(0)$ values of 0.08 and 0.10, respectively.
By leveraging carefully designed evanescently-coupled single-mode photonic resonant cavities, the inherently broad emission spectra of CNTs are significantly narrowed to just a few nanometers. Although challenges remain, including emission instability and limited deposition yield, our results clearly establish the viability of integrating CNT emitters into photonic circuits. This represents a key step forward in the development of room-temperature quantum photonic platforms based on CNT emitters.

\section*{Acknowledgements}

Work supported by JSPS (KAKENHI JP24KF0092, JP24K17627, JP25K21704 and JP23H00262), JST (ASPIRE JPMJAP2310, CREST JPMJCR25A1) and MEXT (ARIM JPMXP1224UT1073, JPMXP1224UT1069). The authors thank the Advanced Manufacturing Support Team at RIKEN for technical assistance. C.F.F. is supported by the RIKEN Special Postdoctoral Researcher Program. F.L.S. and J.Z. acknowledge financial support by the Deutsche Forschungsgemeinschaft (DFG, German Research Foundation) via SFB 1225/3 (Isoquant).

\clearpage

\bibliography{sample}
\clearpage

\setcounter{figure}{0}      
\makeatletter
\renewcommand*\thefigure{S-\arabic{figure}}
\setcounter{section}{0} 
\renewcommand*\thesection{S-\Roman{section}}
\makeatother


\section{Material autofluorescence comparison}

Micro-photoluminescence measurements are performed on three PIC platforms: silicon-rich silicon nitride, stoichiometric silicon nitride, and lithium niobate. In all cases, the films under investigation are 300 nm thick, and are deposited on silicon substrates with buried silicon dioxide layers of 3 µm, 2 µm, and 5.7 µm thickness, respectively. The samples are excited in free space using a titanium-sapphire laser operating at 780 nm with an incident power of 87 µW. Emitted photoluminescence is collected and spectrally analyzed, with integration times of either 5 or 10 seconds.

\begin{figure*}[t]
    \includegraphics[width=0.48\textwidth]{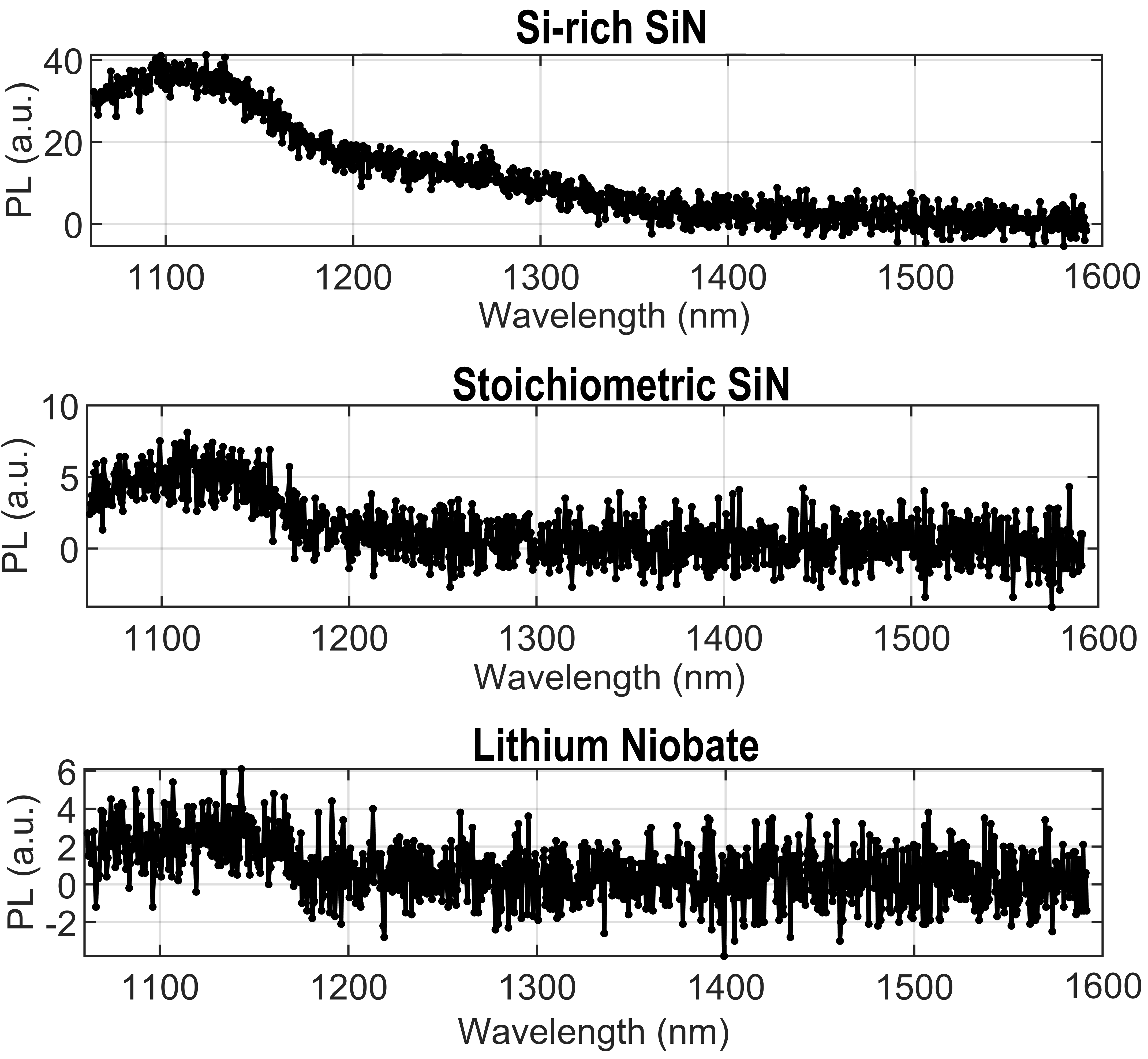}
    \caption{Material near infrared photoluminescence of silicon-rich silicon nitride, stoichiometric silicon nitride and lithium niobate for 87 $\mu$W continuous excitation at 780nm for 5, 10 and 10 seconds respectively.}
    \label{fig:PLmaterial}
\end{figure*}

As shown in Figure \ref{fig:PLmaterial}, the lithium niobate platform exhibits the lowest autofluorescence across the entire measured spectral range, with emission levels approaching the noise floor near the 1300 nm region of interest. It should be noted, however, that the residual signal may originate from the underlying silicon substrate, located 5.7 µm below the surface.

\section{Layout}
\label{layoutsection}
\begin{figure*}
    \includegraphics[width=0.49\textwidth]{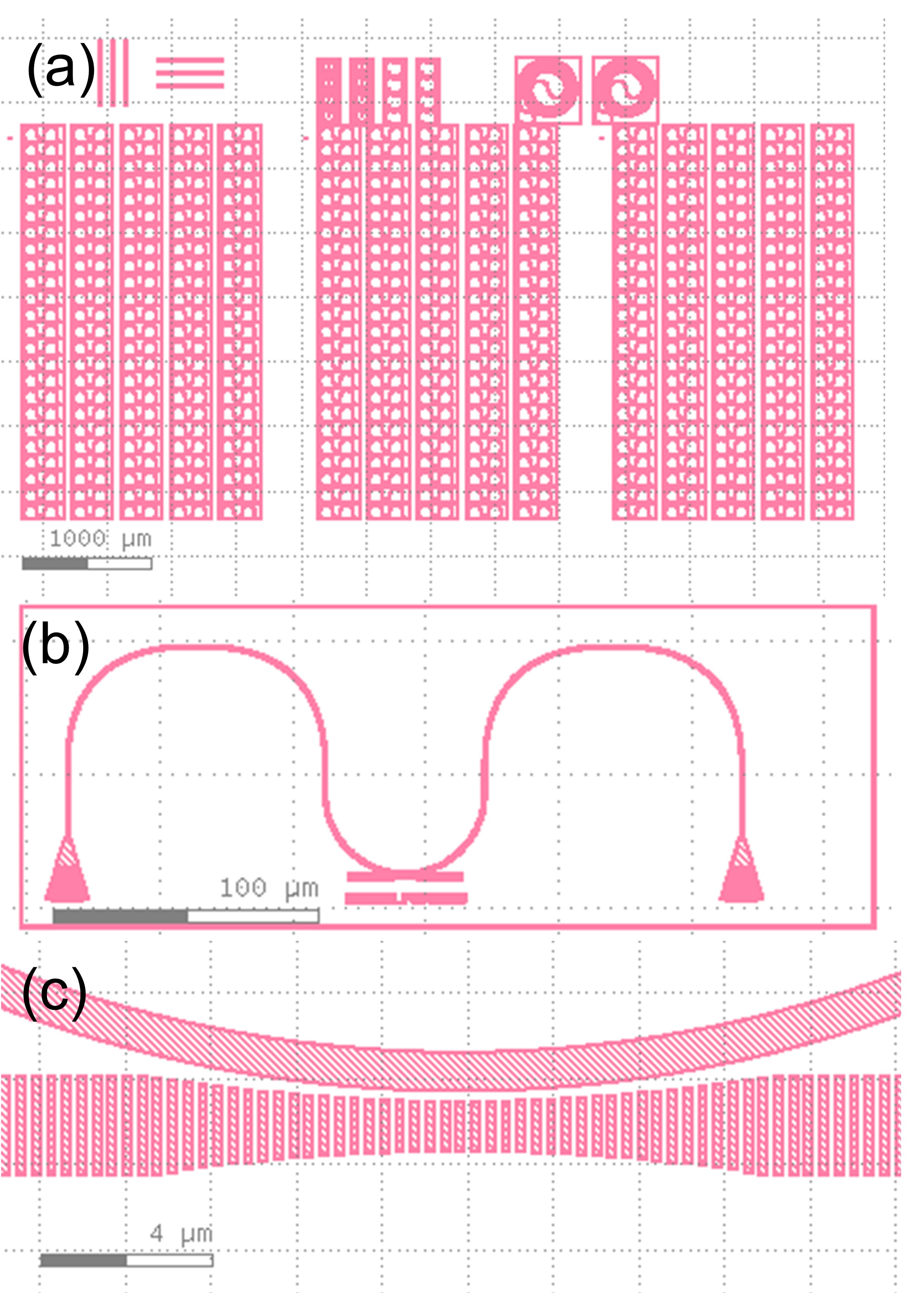}
    \caption{Layout images of (a) cavity array chip, (b) coupling test circuit unit, and (c) coupling cavity.}
    \label{Layoutpics}
\end{figure*}

Grating couplers and waveguides are implemented using the photonic component library developed by Lukas Chrostowski [2]. As depicted in Figure~\ref{Layoutpics}, a large array of cavities is patterned and framed to mitigate local stress and prevent cracks during fabrication. The central region of each cavity tapers in width, thereby shifting the central reflection wavelength and forming a photonic cavity.

\section{Characterization setup}
\label{setupsectionsupporting}
\begin{figure*}
    \includegraphics[width=0.49\textwidth]{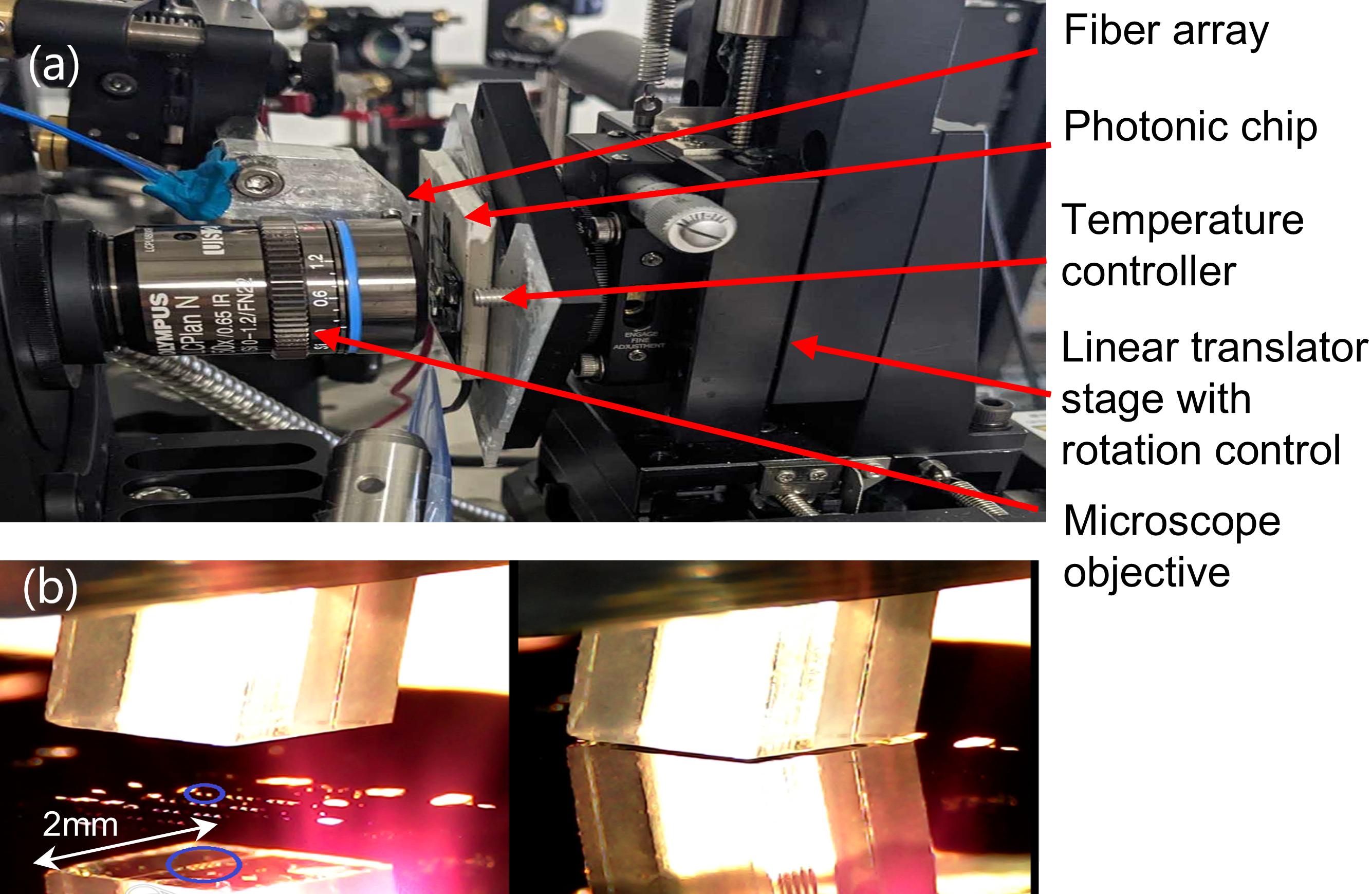}
    \caption{(a) Photoluminescence measurement and PIC coupling setup, (b) and picture of fiber array coupling taken with zoomed camera.}
    \label{fig:setuppic}
\end{figure*}

The characterization setup, illustrated in Figure~\ref{fig:setuppic}, consists of a microscope objective, a fiber array, and a photonic chip stage equipped with linear translation and rotation stages as well as a temperature controller. This configuration maintains a fixed distance between the focal point of the microscope objective and the fiber array channels, enabling seamless switching between free-space excitation and grating coupler-based fiber coupling.

\section{Stamping process example}
\label{sectionstampingimage}
\begin{figure*}
\includegraphics[width=0.49\textwidth]{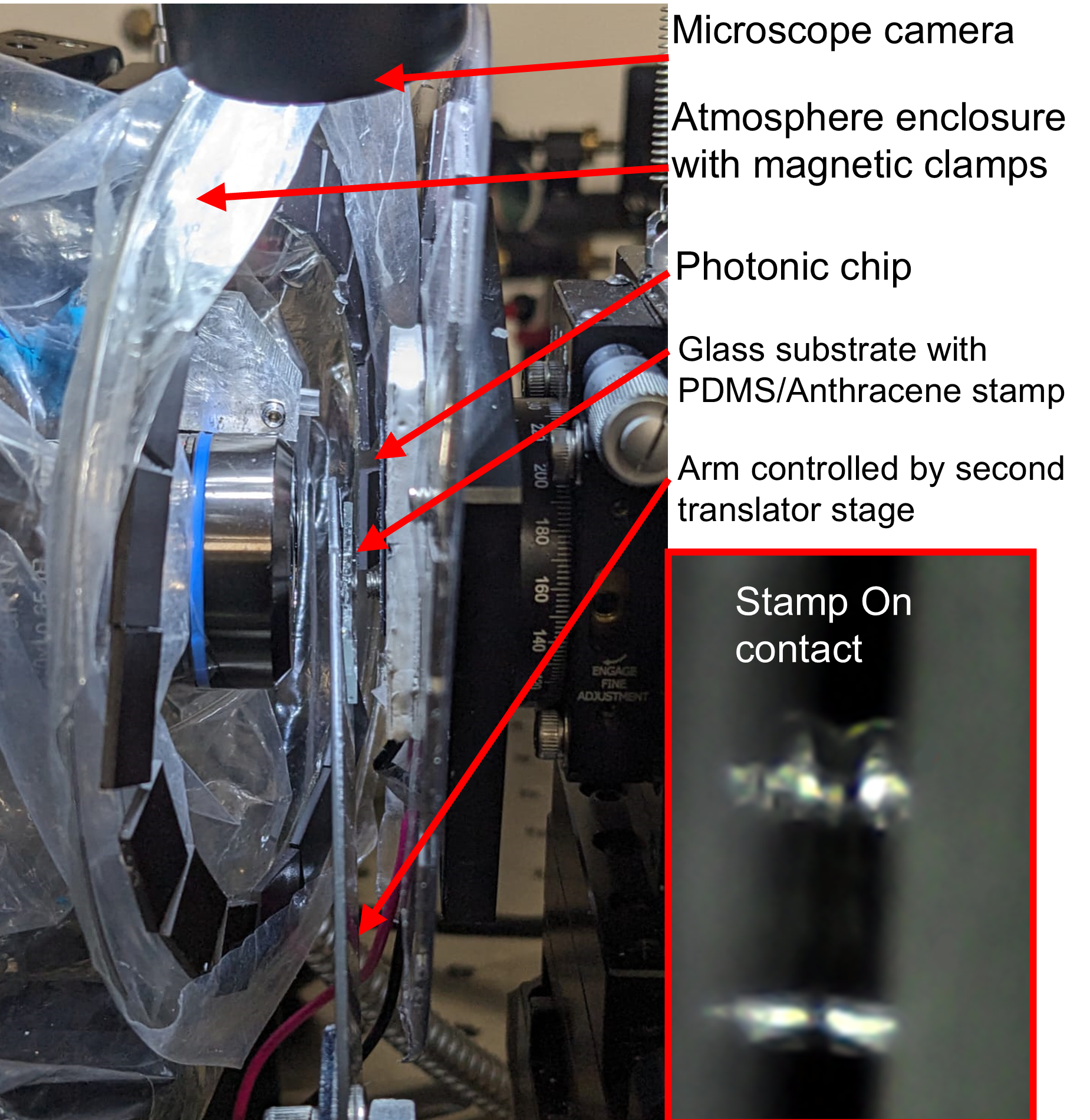}
    \caption{Stamping setup image and zoomed picture of the stamp on contact.}
    \label{setupstamp}
\end{figure*}

\begin{figure*}
    \includegraphics[width=0.49\textwidth]{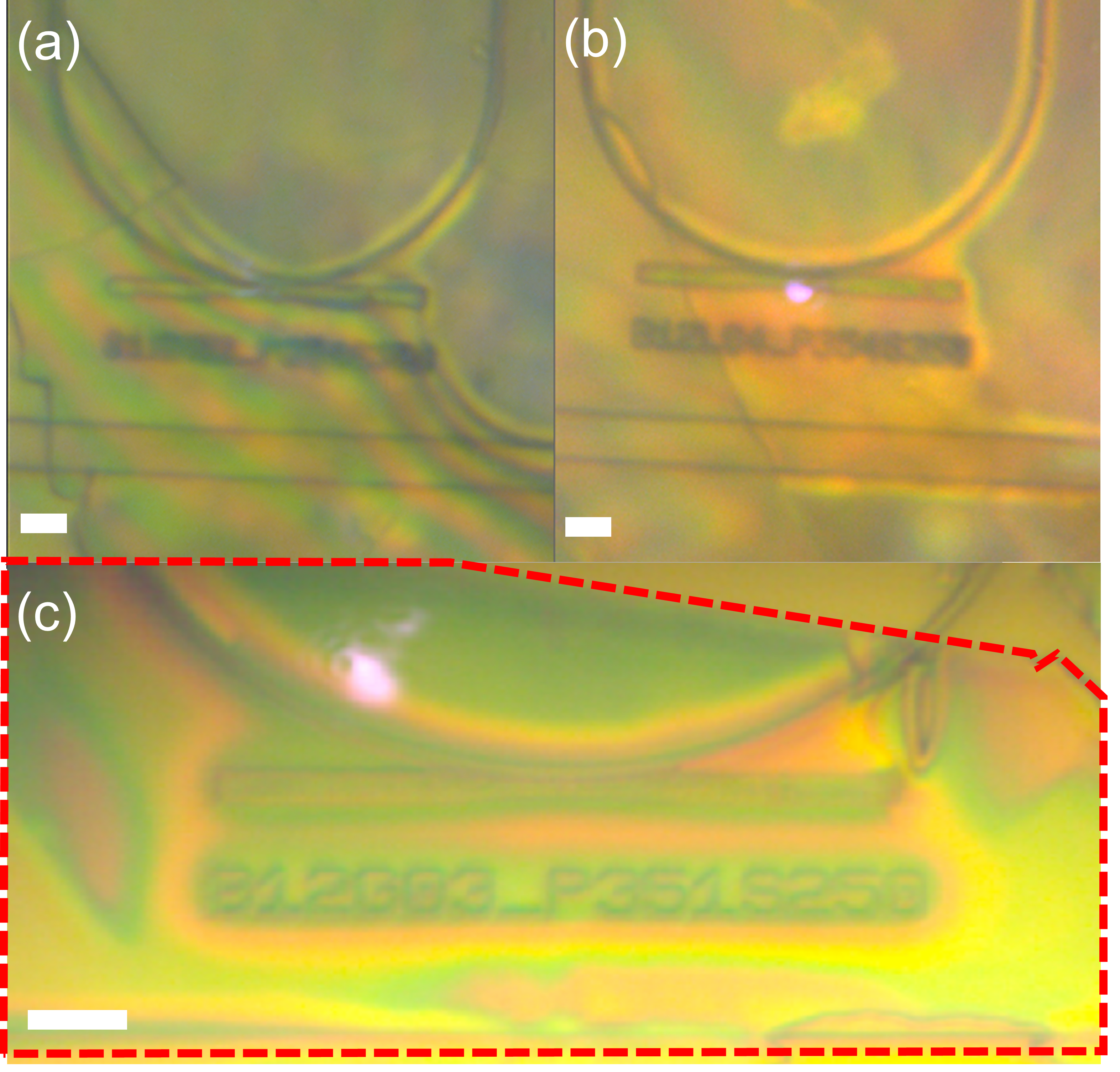}
    \caption{Microscope images of a cavity (a) during stamping process, (b) after stamping, and (c) after PDMS removal. The scale bars are 5~$\mu$m.}
    \label{fig:stampingcav}
\end{figure*}

\begin{figure*}
    \includegraphics[width=0.49\textwidth]{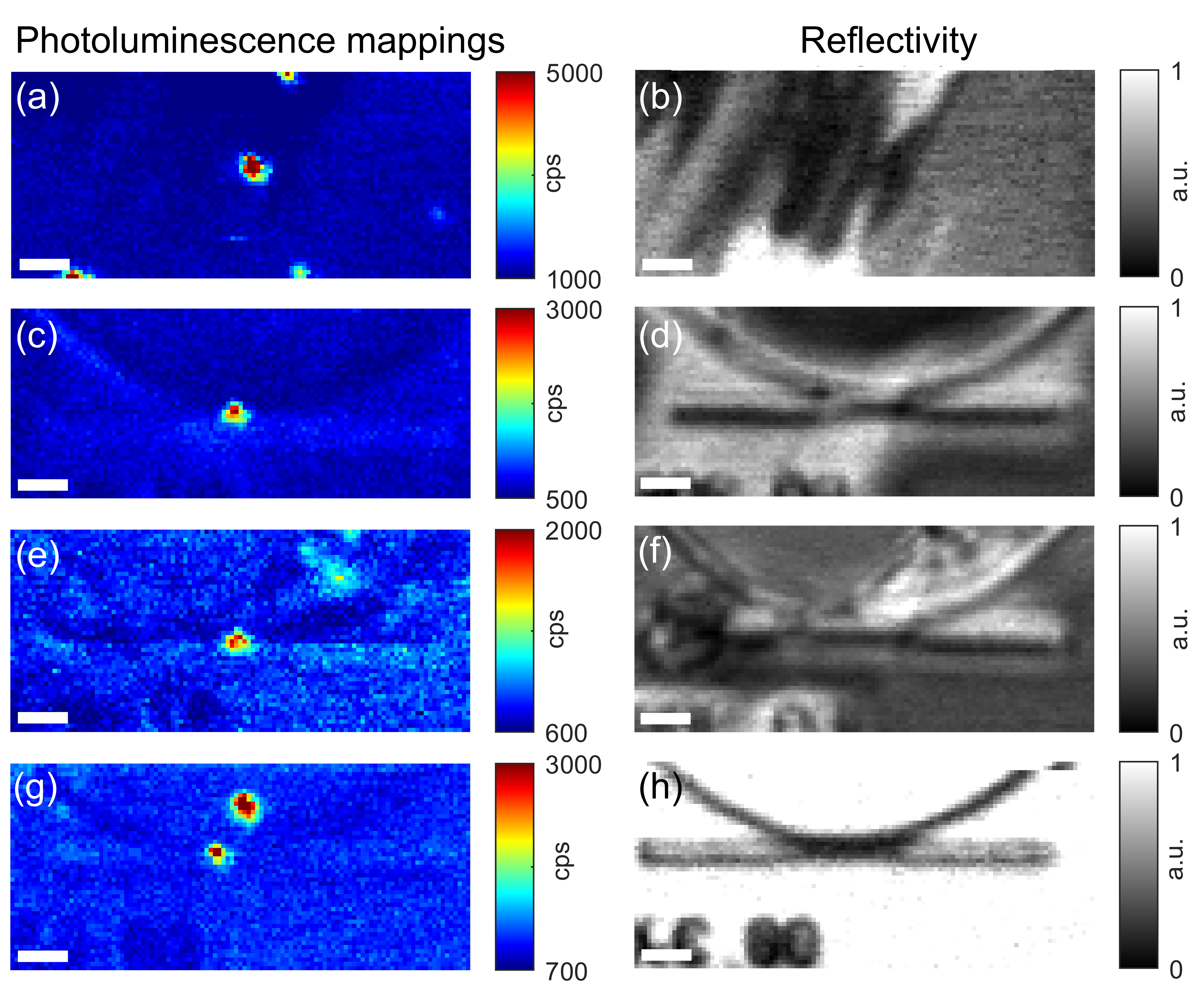}
    \caption{Photoluminescence mappings and associated reflectivity of (a)-(b) CNT-coated anthracene flake and (c)-(d),(e)-(f),(g)-(h) stamped cavities. The scale bars are 5~$\mu$m.}
    \label{fig:stampedPLtopo}
\end{figure*}

This section expands on the CNT stamping procedure described in the main text by providing additional photographs of the experimental setup, the stamping process, and the resulting structures. Figure~\ref{setupstamp} presents the stamping setup with integrated photoluminescence monitoring, along with a side view of the chip being contacted by the small PDMS sheet. Figure~\ref{fig:stampingcav} shows microscope images taken during stamping, highlighting interference fringes around the cavity at the moment of contact and flake fragmentation upon stamp removal, illustrating that the anthracene flake is generally only partially transferred. Finally, Figure~\ref{fig:stampedPLtopo} displays PL and reflection mappings of isolated CNTs on anthracene flakes, as well as after cavity stamping.


\section{Additional cavity-coupled CNT PL spectra}
\label{sectionupportingcavitycoupledspectra}
\begin{figure*}
    \includegraphics[width=0.49\textwidth]{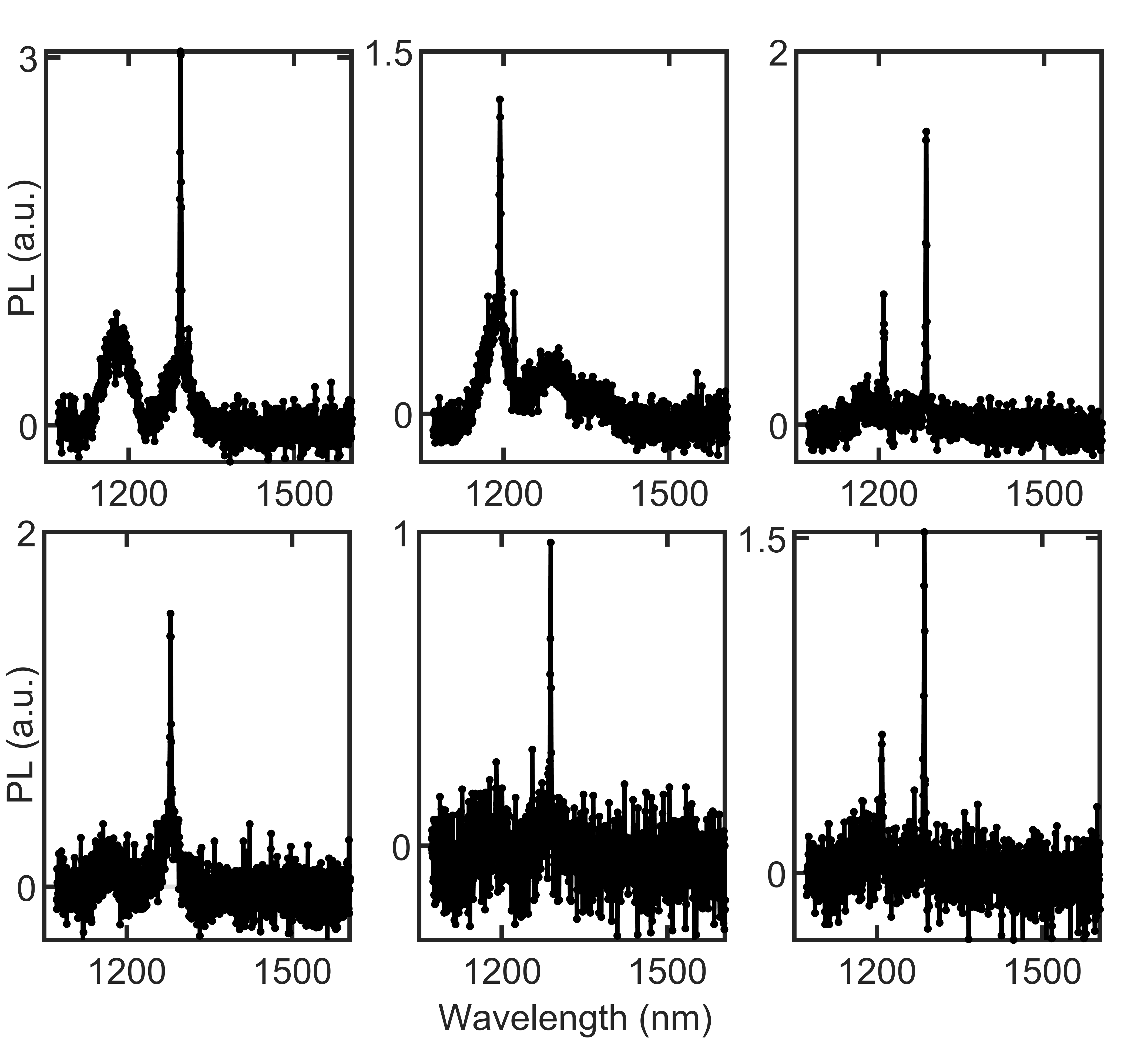}
    \caption{Photoluminescence spectra of several coupled CNTs at 780~nm 1.5~$\mu$W excitation.}
    \label{fig:addinfosPLspectras}
\end{figure*}

Figure~\ref{fig:addinfosPLspectras}  exhibits several examples of coupled CNT spectras as measured after waveguide propagation with fiber arrays. Note that autocorrelation measurements have not necessarily shown single photon propagation for each of these samples. It can also be observed that some cavities with different parameters exhibit one or multiple resonances at distinct wavelengths.

\section{Functionalized CNT stability issues}
\label{sectionsupportingstability}
\begin{figure*}
\includegraphics[width=0.49\textwidth]{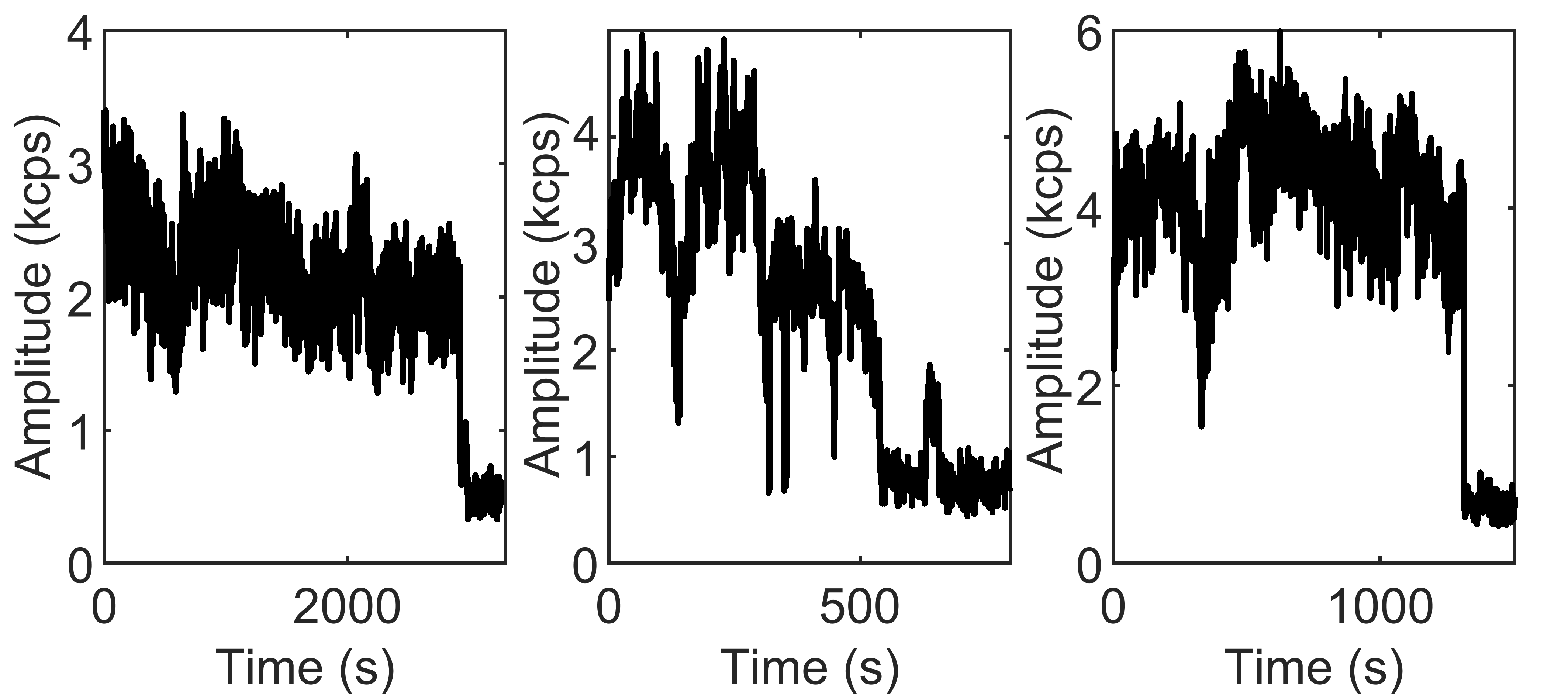}
    \caption{Emission intensities of circuit-coupled CNTs over time.}
    \label{fig:stability}
\end{figure*}

Although CNTs are generally considered chemically stable, solution-processed samples often exhibit blinking behavior and can permanently photobleach after minutes to a few hours of continuous excitation. Representative examples of such blinking dynamics are shown in Figure~\ref{fig:stability}. Interestingly, in some cases, coupled CNTs could be temporarily revived using strong pulsed excitation for a few seconds. However, a gradual decrease in brightness, likely due to defect formation or humidity-induced degradation, ultimately limits the duration over which photon autocorrelation measurements can be performed. Several strategies to mitigate quenching effects are currently under investigation. As illustrated in Figure~\ref{setupstamp}, the setup is upgraded with nitrogen atmosphere control to potentially extend CNT lifetime, but no significant improvement has been observed thus far.

\section{Emission lifetime measurements}
\label{sectionsupportinglifetime}
\begin{figure*}
    \includegraphics[width=0.49\textwidth]{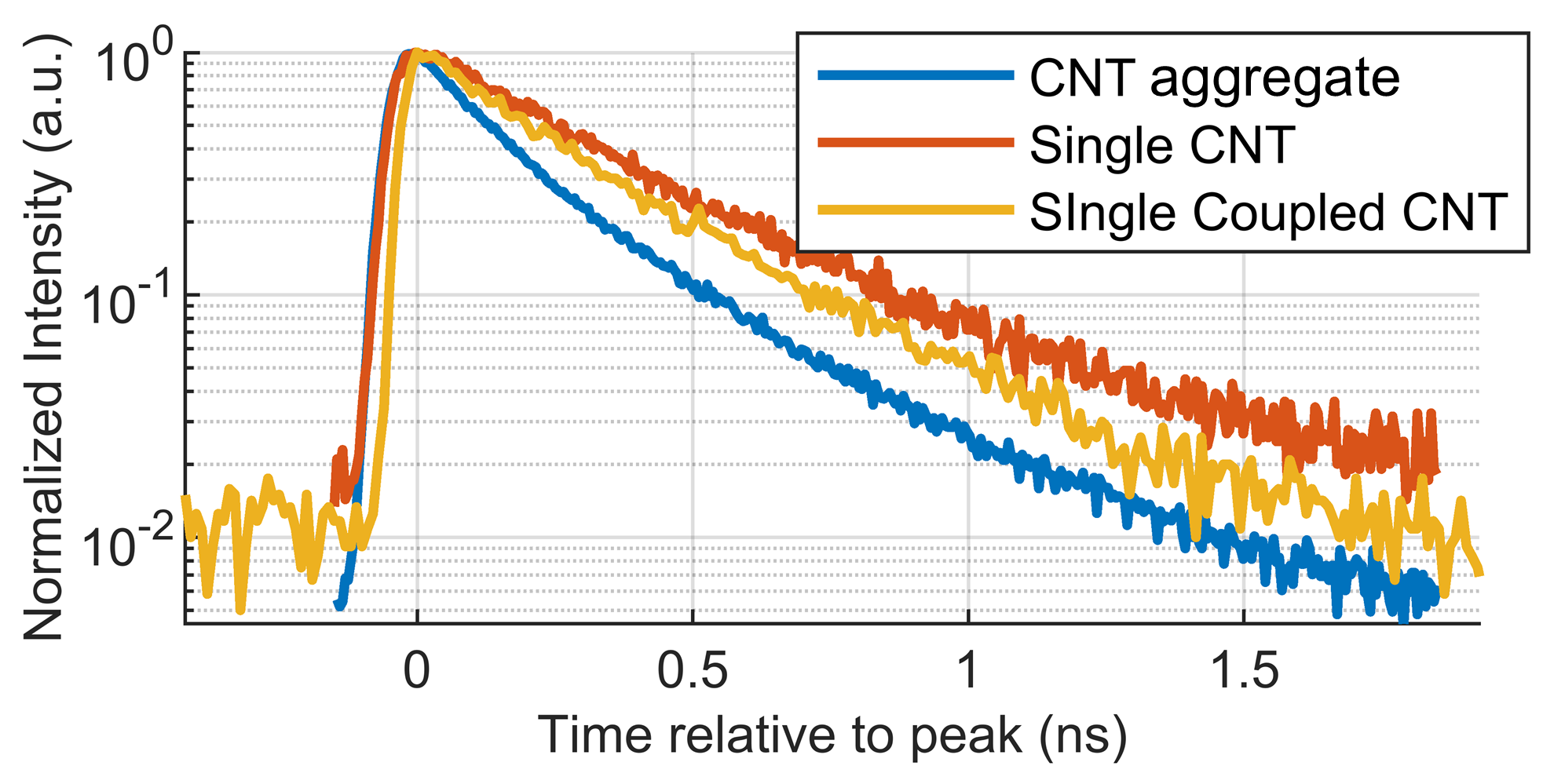}
    \caption{Typical emission lifetime measurements of a CNT aggregate (blue curve), a single CNT (orange curve), and a cavity-coupled CNT (yellow curve).}
    \label{fig:lifetime}
\end{figure*}

Emission lifetime measurements are performed using both free-space excitation and our FA configuration to evaluate potential modifications due to the Purcell effect in cases of cavity coupling. However, as shown in Figure~\ref{fig:lifetime}, no significant lifetime reduction is observed.


\end{document}